# Towards Delay-Tolerant Flexible Data Access Control for Smart Grid with Renewable Energy Resources

Zhitao Guan, *Member, IEEE*, Jing Li, Liehuang Zhu, *Member, IEEE*, Zijian Zhang, *Member, IEEE*
Xiaojiang Du, *Senior Member, IEEE*, Mohsen Guizani, *Fellow, IEEE*

*Abstract* —In the Smart Grid with Renewable Energy Resources (RERs), the Residential Units (RUs) with Distributed Energy Resources (DERs) are considered to be both power consumers and suppliers. Specifically, RUs with excessive renewable generations can trade with the utility in deficit of power supplies for mutual benefits. It causes two challenging issues. First, the trading data of RUs is quite sensitive, which should be only accessed by authorized users with fine-grained policies. Second, the behaviors of the RUs to generate trading data are spontaneous and unpredictable, then the problem is how to guarantee system efficiency and delay tolerance simultaneously. In this paper, we propose a delay-tolerant flexible data access control scheme based on Key Policy Attribute Based Encryption (KP-ABE) for Smart Grid with Renewable Energy Resources (RERs). We adopt the secret sharing scheme (SSS) to realize a flexible access control with encryption delay tolerance. Furthermore, there is no central trusted server to perform the encryption/decryption. We reduce the computation cost on RUs and operators via a semi-trusted model. The analysis shows that the proposed scheme can meet the data security requirement of the Smart Grid with RERs, and it also has less cost compared with other popular models.

## I. Introduction

Energy crisis and environmental pollution have become two critical concerns during the last decade. Smart Grid with Renewable Energy Resources (RERs) is a promising approach to tackle these two problems [1]. RERs are inexhaustible, including light, wind, vibration, heat, biofuel, biomass, and tides. Encouraging and supporting the use of renewable energy power generation technologies, can substantially reduce standard high energy and carbon emission. In the Smart Grid, a two-way communication between the utility and the customers is achieved with the support of information and communication technologies (ICTs) [2]. As shown in Fig.1, in Smart Grid with RERs, renewable distributed energy resources (DERs) at the consumer side will become an important part of power generation. With the consideration of DERs' advantages and benefits, Residential Units (RUs) that are equipped with DERs will supplement their daily electricity demands and assist in reducing the pressure on the main grid during peak hours. If there is any electricity generation excess, RUs also can sell the surplus of clean energy to the main grid for a certain income by acting as small-scale electricity suppliers (SESs) [3].

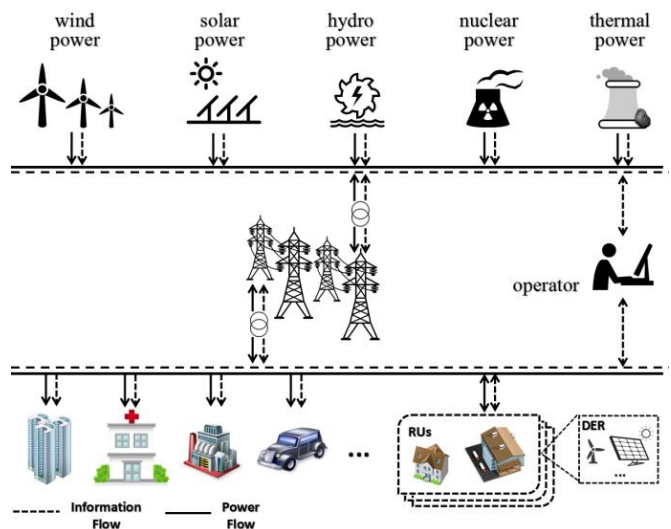

Fig.1. The architecture of Smart Grid with RERs

Those RUs will send their price and bid information to the power network. If their prices are competitive enough, the network operator will decide to trade with them. Therefore, it is important to ensure that the individual price is unknown to each other, to avoid a malicious low-price competition. Thus, the Smart Grid is expected to provide security to the transmitted information. As far as we know, Attribute Based

● The work is partially supported by the National Natural Science Foundation of China under Grant 61402171, the Fundamental Research Funds for the Central Universities under grant 2015ZD12, grant 2016MS29, as well as the US Air Force Research Lab under grant AF16-AT10 and the Qatar National Research Fund under grant NPRP 8-408-2-172.

● Liehuang Zhu is the Corresponding author (liehuangz@bit.edu.cn).

● Zhitao Guan and Jing Li are with School of Control and Computer Engineering, North China Electric Power University, Beijing, China. (E-mail: guan@ncepu.edu.cn, li_jing@ncepu.edu.cn).

● Liehuang Zhu and Zijian Zhang are with School of Computer, Beijing Institute of Technology, Beijing, China. (E-mail: liehuangz@bit.edu.cn, zhangzijian@bit.edu.cn).

● Xiaojiang Du is with Department of Computer and Information Science, Temple University, Philadelphia, PA USA. (E-mail: dxj@ieee.org).

● Mohsen Guizani is with University of Idaho, Moscow, ID USA. (E-mail: mguizani@ieee.org).

Encryption (ABE) [4] has been already used in various practical scenarios in order to achieve the data security and realize a fine-grained access control.

In this paper, the information from the same area is considered to share similar attributes, and the agency operators (AOs) are allowed to have the authority to specify access policies. Therefore, the Key Policy Attribute Based Encryption (KP-ABE) [5], an important variant of ABE, is a better choice. Actually, these RUs with DERs that share the same attributes may belong to the same micro-grid, which is not the focus of this paper. Therefore, this is not addressed in this paper.

When the AOs require the bid information to be checked, there are typically several methods to choose from. One of the intuitive methods is to set a central trusted server to collect all of the bid information from the corresponding bidders. Generally, as the RUs' trading behaviors are difficult to predict, the central trusted server will set a transaction period, during which any received information is considered valid and any information that misses the deadline will be deemed void. The drawback of this is that the AOs can only request decryption after this period is over. Alternatively, we can abandon the central trusted server, making it possible that each set of RUs can independently encrypt its information. This method greatly reduces the waiting time of the AOs. However, it will result in frequent decryptions.

Both of the above methods will cause varying degrees of computation overhead. To solve this problem, we propose a flexible access control scheme with delay tolerance. The main contributions in this paper can be summarized as follows:

- We propose a scheme that allows some of the RUs that have generated information in a short time to first encrypt and upload, while other RUs' operations can be delayed. This can reduce the AOs' waiting time for decryption. As a result, all of these ciphertexts can be obtained indiscriminately by AOs and decrypted once rather than multiple times.
- By combing the KP-ABE and secret sharing scheme (SSS), we achieve a secure and flexible access control scheme with delay tolerance. In this paper, there is no central trusted server to participate in encryption/decryption, we replace the trusted ones with the semi-trusted. They assist with calculation and reducing the computation cost on RUs and AOs without impacting the data security.
- We show the security proof and analysis, which demonstrate that the proposed scheme can fulfill the security requirements in a practical scenario. Compared with the traditional schemes, the experimental results indicate that our scheme can effectively reduce the time cost and improve system efficiency.

The rest of this paper is organized as follows. In Section II, related work is introduced. In Section III, the preliminaries are given. In Section IV, the system model and security model are described. In Section V, we present the details of the proposed scheme. In Section VI and VII, its security analysis and performance evaluation are conducted, respectively. Section VIII concludes the paper.

## II. RELATED WORK

With the gradual improvement of Smart Grid construction, energy consumption and environment pollution have attracted attention and become widely studied. Many researchers believe that Smart Grid integrated with the RER is the best way at present. Sims *et al*. presented the carbon emission and mitigation cost comparisons between the traditional energy resources and renewable energy resources [6]. Yu *et al*. focused on several communication technologies available for Smart Grid with RER [7]. Other relevant research details are described in [8-9].

Another important concept in this paper is DER, which means the small power generators typically located at the users' sites. The generated energy can meet the growing customer needs more or less [10]. Due to the development of RER technologies, there is plenty of literature on the combination of DER and RER, such as [11].

No matter whether using the traditional grid or the emerging Smart Grid with RER, data security and user privacy always present some challenging issues. Thus, there have been considerable research efforts to understand the definition and future development of Smart Grid [12]. This research primarily summarizes the current Smart Grid technologies and the key issues by pointing out the important significance of improving system reliability and ensuring data security in Smart Grid. In [13], Xu *et al*. listed the security challenges and analyzed the attack risks, and then proposed a variety of defense strategies. There are some other related work considering the security issues, such as [14-18]. In order to address variety of security issues, many cryptographic schemes have been proposed. Here we only introduce three essential kinds of techniques.

*A. Homomorphic Encryption*

Homomorphic encryption allows arithmetic operations to be performed on ciphertext and gives the same result as if the same arithmetic operation is done on the plaintext [19]. It has been viewed as one of the promising methods to be employed in Smart Grid to provide data security and privacy preserving. He *et al*. introduced an application of public key encryption of Smart Grid in [20]. They adopted the partially homomorphic encryption algorithm to protect the data exchange process between consumers and utilities.

*B. Attribute-Based Encryption (ABE)*

ABE [4] is often applied to the Smart Grid; it could not only provide the data security, but can also achieve the fine-grained data access control. The key features of ABE are as follows: first, the identities of the encrypted data are expressed by several descriptive attributes. Second, as long as the number of matching attributes reach the specified threshold, the ciphertexts can be decrypted.

There are two important variants of ABE: One is KP-ABE that is proposed by Goyal *et al*. in [5], they developed this new cryptosystem based on Sahai's work [4]. In their scheme, the ciphertext is associated with a set of attributes. Users define the access policies, and the attributes in an access policy are organized into a tree structure (described as access tree). The other one is Ciphertext Policy Attribute Based Encryption (CP-ABE) proposed by Bethencourt in [21]. In their work, the data owner constructs the access tree using visitors' identity information. The user can decrypt the ciphertext only if the attributes in their private key match the access tree. Fadlullah *et al.* focused on the applicability of KP-ABE in Smart Grid in [22]. They considered to use KP-ABE to broadcast a single encrypted message to a specific group of users, which ensured the system's efficiency and communication security simultaneously. The latest researches on ABE can be access in [23-25].

*C. Secret sharing scheme (SSS)*

SSS is a hot spot of cryptography; it is used for sharing a secret among a group of parties, each of whom only obtain a piece of the secret (namely a share of the secret). The most basic secret sharing scheme is first proposed by Shamir [26]. In his paper, no single party could infer any information about the secret with its own share. The only way to reconstruct the secret is to combine a certain number of shares. After Shamir's work, there were many practical schemes being proposed to adapt to various scenarios, including [27-30].

Despite the large number of research that focus on a variety of security issues, no one has taken the system efficiency and user delay-tolerance into consideration. In this paper, to improve the flexibility of the system and realize the fine-grained access control, we combine KP-ABE [5] with the scheme proposed by Pedersen [30], which requires all participants to complete the encryption without a trusted third party.

## II. PRELIMINARY

*A. Bilinear Maps*

Let $G_0$ and $G_1$ be two multiplicative cyclic groups of prime order $p$ and $g$ be the generator of $G_0$. The bilinear map $e$ is, $e: G_0 \times G_0 \to G_1$, for all $a, b \in \mathbb{Z}_p$:

- Bilinearity: $\forall u, v \in G_1, e(u^a, v^b) = e(u, v)^{ab}$.
- Non-degeneracy: $e(g, g) \neq 1$.
- Symmetric: $e(g^a, g^b) = e(g, g)^{ab} = e(g^b, g^a)$.

*B. Discrete Logarithm (DL) Problem:*

Let $G$ be a multiplicative cyclic group of prime order $p$ and $g$ be its generator. Given a tuple $<g, g^x>$, where $g \in_R G$ and $x \in \mathbb{Z}_P$ are chosen as input uniformly at random, the DL problem is to recover $x$.

**Definition 1** The DL assumption held in $G$ is that no probabilistic polynomial-time (PPT) algorithm $\mathcal{A}$ can solve the DL problem with negligible advantage. We define the advantage of $\mathcal{A}$ as follows:

$$\Pr[\mathcal{A} < g, g^x >= x]$$

The probability is taken over by the generator $g$, randomly chosen $x$, and the random bits consumed by $\mathcal{A}$.

*C. The Decisional Bilinear Diffie-Hellman (BDH) Assumption*

Let $G$ be a multiplicative cyclic group of prime order $p$ and $g$ be its generator. Let $a, b, c \in \mathbb{Z}_P$ be chosen randomly. Given two tuples, where

$$<A = g^a, B = g^b, C = g^c, e(g, g)^{abc}>,$$
$$<A = g^a, B = g^b, C = g^c, e(g, g)^z>.$$

**Definition 2** The decisional BDH assumption holds only when there is not a probabilistic polynomial-time algorithm $\mathcal{B}$ that is able to distinguish the two above with more than a negligible advantage. The advantage is

$$\left|\Pr[\mathcal{B}(A, B, C, e(g, g)^{abc}) = 0] - \Pr[\mathcal{B}(A, B, C, e(g, g)^z) = 0]\right|$$

where the probability is taken over the random parameters $g$, $a$, $b$, $c$ and the random bits consumed by $\mathcal{B}$.

*D. Symmetric Encryption*

In a symmetric encryption scheme, two probabilistic polynomial time (PPT) algorithms exist: ***Enc(K, m)→C***, which maps a symmetric key $K \in \mathcal{K}$ and a message $m \in \{0,1\}^*$ to the ciphertext $C$, and ***Dec(C, K)→m***, retrieving the message $m$ with symmetric key $K$.

The security game is described as follows: An adversary submits two messages $m_0$ and $m_1$ with the same length $\kappa$ to the challenger. Then, the challenger randomly flips a coin $b$ and encrypted $m_b$ with a symmetric key that $K \in \mathcal{K}$. The ciphertext will be transmitted to the adversary, and he/she gives a guess $b'$ about $b$.

**Definition 3** The one-time symmetric encryption scheme is semantically secure only when for any PPT adversary,

$$Adv_\mathcal{A}(\kappa) = \left|\Pr[\mathcal{A} \text{ wins } b' = b]\right|$$

is negligible in $\kappa$.

*E. Access Structure in Key-policy Attribute Based Encryption*

**Definition 4** Let $P = \{P_1, P_2, ..., P_n\}$ be a set of participants and let $U = 2^{\{P_1, P_2, ..., P_n\}}$ be the universal set. If $\exists AS \subseteq U \setminus \{\varnothing\}$, then $AS$ can be viewed as an access structure. If $A \in AS, \forall B \in U, A \subseteq B$, and $B \in AS$, $AS$ is considered to be a monotonic access structure. The sets in $AS$ are defined as authorized sets, while the other sets are regarded as unauthorized sets.

- *Construction for the access tree*

In a KP-ABE scheme, let $\mathcal{T}$ be the access tree and all of the interior nodes represent the threshold value (including the root node $\mathcal{R}$), such as *AND* (*n* of *n*), *OR* (1 of *n*), and *n* of *m* (*m*>*n*). Set $k_x$ denotes the threshold of node *x*. Define some functions, ***parent***(*x*), denoting the parent node of node *x*, ***index***(*x*) returns the number associated with the node *x*, where the index uniquely assigned to nodes, ***att***(*x*) is used only if *x* is a leaf node and returns the attribute associated with the leaf node $x$ in the tree. At the beginning of the encryption, the algorithm randomly chooses a secret number $s$ and conducts a polynomial for each interior node from top to bottom.

● *Satisfy the tree*

The tree is constructed from top to bottom within the encryption, while the decryption order is reverse.

When some users require ciphertext decryption, they have to start from the leaves nodes. To retrieve the secret, we define the Lagrange coefficient $\Delta_{i,S}$, for $i \in \mathbb{Z}_p$, and for $\forall x \in S$,

$$\Delta_{i,S(x)} = \prod_{j \in S, j \neq i} \frac{x-j}{i-j}.$$

*F. (t, n) Threshold Secret Sharing*

The most basic secret sharing scheme is the $(t, n)$ threshold scheme, which was first proposed by Shamir [26]. In this paper, we adopt the scheme proposed by Pedersen [30]. Assume that there are $n$ parties: $P_1, P_2,..., P_n$ in the system and the threshold is $t$. All of the parties select a unique random number $x_i$ as the individual identity and broadcast to the public. Each party $P_i$ chooses a number $s_i$ at random as his personal secret, and thus the total secret can be obtained by computing $s = \sum s_i$. Each party $P_i$ constructs a polynomial $f_i(x)$ for the other parties, which has a degree of $t$-1 and $f_i(0) = s_i$. These polynomials are distributed to other $n$-1 parties independently. Additionally, the personal secret $s_i$ is secure as it is polynomially indistinguishable from the uniform distribution.

## IV. SYSTEM MODEL AND SECURITY MODEL

*A. System Model*

Consider that in certain areas, there may be large number of smart communities. Each of them involves a large number of RUs. Generally, the generated information of those RUs in the same community share the same attributes. We set a semi-trusted central aggregator (**CA**) in charge of assisting encryption. All of the encrypted bid information is sent to the network via CA. It first collects all the data from RUs. When the number of RUs reaches the threshold, it will perform calculation on them and send them to the network. The delayed encrypted data will be sent to the network via CA in chronological order. When it comes to decryption, the agency operators (**AO**) will deal with the trading information via the central dispatcher (**CD**). AOs can construct the access structures that consist of the attributes of the specified bid information. As soon as there is an operator that requires the decryption of ciphertexts, CD performs the pre-decryption beforehand.

*B. Security Assumption*

In this paper, the security assumption of the entities is defined as follows. The CA and CD are always online and assist with encryption/decryption. Both of them are considered to be "honest-but-curious". Neither can collude with the RUs, because in a real scenario, any of the RUs can be the bid information owner and the collusion may threaten the individual benefit. The RUs generate and encrypt their own bid information to trade the electricity, however, they tend to be curious about others' information, as it is highly beneficial. In addition, there may be some unauthorized operators colluding to obtain illegal access authority. For instance, if operator A has the key associated with the access structure "X *AND* Y", and operator B has the key associated with the access structure "Y *AND* Z", it is illegal for them to collude to decrypt a ciphertext whose only attribute is Y.

*C. Security Model*

Here, we introduce the universal security model of our system, which is defined similar to Goyal's scheme in [5]. In this security model, there is an adversary $\mathcal{A}$ and a challenger $\mathcal{C}$. The adversary is allowed to select a set of attributes and challenge on an encryption to the access structure *AS\** and query for any secret key SK. The challenger is responsible for the ciphertext generation under the *AS\** and the secret key generation. The security game is described as follows:

**Initial:** The adversary $\mathcal{A}$ first selects a set of attributes $\gamma$ and sends it to the challenger $\mathcal{C}$.

**Setup:** The challenger $\mathcal{C}$ runs this Setup algorithm to generate all of the public keys and master keys.

**Phase 1** The adversary $\mathcal{A}$ issues queries for repeated private keys for various access structure $AS_i$. If for $\forall i, \gamma \in AS_i$, then the queries are aborted. Else, $\mathcal{C}$ generates the corresponding secret keys for $\mathcal{A}$.

**Challenge** The adversary submits two messages with equal length, $M_0$ and $M_1$. The challenger $\mathcal{C}$ randomly flips a coin $b$, and encrypts $M_b$ with the selected set $\gamma$. Then, the generated ciphertext *CT\** will be given to $\mathcal{A}$.

**Phase 2:** Repeat **Phase 1**.

**Guess:** The adversary $\mathcal{A}$ outputs its guess $b' \in \{0,1\}$ for $b$ and wins the game if and only if $b' = b$.

The advantage of an adversary $\mathcal{A}$ in this game is defined as,

$$Adv(\mathcal{A}) = \left| \Pr[b'=b] - \frac{1}{2} \right|,$$

where the probability is taken over the random bits used by the challenger and the adversary.

**Definition 5** A KP-ABE scheme is secure in the Selective-Set model of security if all polynomial time adversaries have at most a negligible advantage in the Selective-Set game.

**Definition 6** A KP-ABE scheme is CPA-secure if challengers allow for decryption queries in Phase 1 and Phase 2.

## III. DETAILS OF THE PROPOSED SCHEME

*A. Overview*

In this paper, the overview of our system is shown in Fig. 2. The CA will be based on the actual situation to develop a transaction period. Only the information submitted within this period shall be deemed valid, otherwise, it will be considered invalid. The threshold value that determines the number of RUs uploading their information can be specified by the system operators in advance according to the actual needs. Once the number of trading RUs reaches the threshold value, they can begin to encrypt and upload. Then, the CA will be ready for receiving the semi-finished ciphertext from the RUs and converting them into finished ones. However, the CA is not involved in the encryption of plaintexts. As long as the AO sends his/her request for ciphertexts, the CD obtains the desired ciphertexts and pre-decrypts for the AO without detecting any plaintexts information.

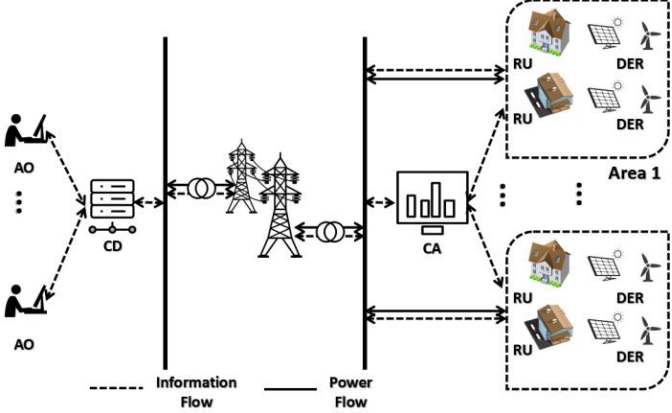

Fig.2. System Model

*B. Algorithms*

Here we will give the details of the algorithms in our system.

*1) Setup*

This setup algorithm will choose a bilinear group $G_0$ of prime order $p$ with generator $g$. Assume that the universe of attributes in the system: $\mathcal{U} = \{1, 2, ..., n\}$. For $\forall i \in \mathcal{U}$, the algorithm will associate this attribute with a unique element $t_i$ in $\mathbb{Z}_p^*$ and choose two random exponents: $y, \alpha \in \mathbb{Z}_p$. The public key and the master key are published as:

$$PK = \left\{ T_1 = g^{t_1}, ..., T_{|\mathcal{U}|} = g^{t_{|\mathcal{U}|}}, Y = e(g,g)^y, A = g^\alpha \right\}, \quad (1)$$

$$MK = \left\{ t_1, t_2 ... t_{|\mathcal{U}|}, y, \alpha \right\}. \quad (2)$$

*2) Key Generation*

This operation is implemented by a specific trusted server independently. It is responsible for registering legal AOs, authorizing the access policies, and generating the secret keys SKs accordingly.

This algorithm will construct an access tree for the authorized access structure AS. It takes the access structure and the master key as inputs, and outputs a private key for the AO. The ciphertexts can only be decrypted by the key owner if their attributes satisfy the tree. All of the steps are similar with that of [5]. After constructing the access tree, for each leaf node $x$, the key components are computed as follows:

$$D_x = g^{\frac{q_x(0)}{t_i}}, \text{where } i = att(x). \quad (3)$$

Thus, the user's decryption key is, $D : \{ \mathcal{T}, D_x \}$.

*3) Encryption*

The RUs' bid information should be transmitted over the network securely. The best situation is that all of the potential trading RUs generate their own bid information, which can be packed and encrypted together as one ciphertext. However, in a real scenario, it is possible for there to be only a few RUs generating their bid information over a period of time. In this paper, we allow the latency of some RUs' encryption.

The encryption operation is performed by the CA and several RUs that belong to the same community. First, the CA will set a transaction period. In this period of time, $k$ RUs ($0 < k \leq n$, $k$ denotes the threshold value that can be specified by the system in advance, $n$ denotes the total number of RUs in the current area.) have generated their data, and the CA will first inform all those RUs to generate a polynomial $f_i(x)$ by themselves, which the degree is $k$-1. Each of the $RU$s will choose at random a number $s_i \in \mathbb{Z}_p$ and set

$$f_i(0) = s_i, \quad (4)$$

$$s = \sum_{i=1}^{n} s_i = \sum_{i=1}^{n} f_i(0). \quad (5)$$

Additionally, all of the RUs in the same area will obtain a symmetric encryption key $\mathcal{K}$ from a specific trusted server that is totally unknown to the CA.

$RU_i$ computes with $\mathcal{K}$ as follows:

$$Y' = g^{s_i}. \quad (6)$$

Then set the ciphertext id as $CID_i$:

$$CID_i = Enc_{\mathcal{K}}(Y'). \quad (7)$$

Then, it will broadcast its $CID_i$ to the other $n$-1 RUs. As long as it obtains others' $CID$, for $\forall 1 \leq j \leq n, j \neq i$, computes $f_i(CID_j)$ and sends it to $RU_j$.

When $RU_i$ gets all the $f_j(CID_i)$ from the other RUs ($\forall 1 \leq j \leq n, j \neq i,$), it will set a new function:

$$h_i(CID_i) = f_1(CID_i) + ... f_i(CID_i) + ... f_n(CID_i). \quad (8)$$

Then, the algorithm will encrypt the RU's bid information. For the attributes set $\gamma$, computing as follows:

$$\hat{C}_i = M_i e(g^\alpha, g^{s_i}) = M_i e(g,g)^{\alpha s_i}, \quad (9)$$

$$\tilde{C}_i = e(g,g)^{y h_i(CID_i)}, \quad (10)$$

$$C_{u,i} = T_u^{h_i(CID_i)} \quad u \in \gamma. \quad (11)$$

Thus, the semi-finished ciphertext is:

$$C_i' = \left\langle \gamma, \hat{C}_i, \tilde{C}_i, \{C_{u,i}\}_{u \in \gamma} \right\rangle. \quad (12)$$

All the $k$ semi-finished ciphertexts from the RUs will be transmitted to the CA. Once it obtains $k$ ciphertexts, it performs the polynomial interpolation as follows:

$$e(g,g)^{ys} = \prod_{i=1}^{k} e(g,g)^{yh_i(CID_i) \prod_{j=1,j\neq i}^{k} \frac{CID_j}{CID_j - CID_i}} \quad (13)$$

$$= e(g,g)^{y \sum_{i=1}^{k}(h_i(CID_i) \prod_{j=1,j\neq i}^{k} \frac{CID_j}{CID_j - CID_i})}.$$

The ciphertexts are changed to $C_i = \langle \gamma, \hat{C}_i, \{C_{i,t}\}_{t \in \gamma} \rangle$ where

$$\hat{C}_i = M_i e(g,g)^{\alpha s_i} e(g,g)^{ys}. \quad (14)$$

They will be transmitted to the network. In the near future, there may be other RUs generating information in succession, performing the same operation as above, and submitting the semi-finished ciphertexts to the CA. While the CA does not need to do anything except that: $\hat{C}_i = M_i e(g,g)^{\alpha s_i} e(g,g)^{ys}$. As $e(g,g)^{ys}$ is known.

*4) Decryption*

When the AO requires to view the bid information of some certain area, they tend to obtain and access all the *RU*s' information at once. First, he/she will submit the request for decryption and get $l$ ($k \leq l \leq n$) ciphertexts via CD. It will pre-decrypt the ciphertexts by querying the symmetric key of the area. For $\forall 1 \leq i \leq m$,

$$Y' = Dec_\mathcal{K}(CID_i) = g^{s_i}, \quad (15)$$

$$\frac{\hat{C}_i}{e(g^\alpha, g^{s_i})} = \frac{M_i e(g,g)^{\alpha s_i} e(g,g)^{ys}}{e(g,g)^{\alpha s_i}} = M_i e(g,g)^{ys}. \quad (16)$$

The ciphertexts are changed to $\tilde{C}_i = \langle \gamma, \breve{C}_i, \{C_{u,i}\}_{u \in \gamma} \rangle$ where

$$\breve{C}_i = M_i e(g,g)^{ys}. \quad (17)$$

Thus, as soon as he/she gains $l$ ($k \leq l \leq n$) ciphertexts from the network, he/she selects $k$ of them and computes as follows:

$$T_u^s = \prod_{i=1}^{k} T_u^{h_i(CID_i) \prod_{j=1,j\neq i}^{k} \frac{CID_j}{CID_j - CID_i}}$$

$$= T_u^{\sum_{i=1}^{k}(h_i(CID_i) \prod_{j=1,j\neq i}^{k} \frac{CID_j}{CID_j - CID_i})}. \quad (18)$$

After that, the AO decrypts the ciphertexts as the authors did in [5]. First, for $\forall i \in \gamma$,

$$e(g^{\frac{q_z(0)}{t_u}}, g^{t_u s}) = e(g,g)^{q_z(0)s}, \quad (19)$$

$$F_x = \prod_{z \in S_x} F_z^{\Delta_{i,S'_x}(0)}, \text{ where } i = index(z), S'_x = \{index(z) : z \in S_x\}$$

$$= \prod_{z \in S_x} (e(g,g)^{q_z(0)s})^{\Delta_{i,S'_x}(0)}$$

$$= \prod_{z \in S_x} (e(g,g)^{sq_{parent(z)}(index(z))})^{\Delta_{i,S'_x}(0)} \quad (20)$$

$$= \prod_{z \in S_x} e(g,g)^{sq_x(i)\Delta_{i,S'_x}(0)}$$

$$= e(g,g)^{sq_x(0)}.$$

For root node R,

$$F_R = e(g,g)^{ys}. \quad (21)$$

By this method, the message can be decrypted together at one-time:

$$M_{i(1 \leq i \leq m)} = \frac{M_i e(g,g)^{ys}}{e(g,g)^{ys}} \quad (22)$$

IV. SECURITY ANALYSIS

We analyze the security properties of our system considering the security model defined in this Section.

*A. System Security*

We prove the security of our scheme according to the security model defined in Section III.

**Theorem 1**. When the Decisional BDH assumption holds, there is no adversary that can break our scheme in the Attribute-based Selective-Set Model.

**Proof:** We prove this theorem by the following game. Suppose an adversary $\mathcal{A}$ can attack our scheme in Selective-Set model with non-negligible advantage.

Suppose there is a Challenger $\mathcal{C}$ that can play the Decisional BDH game with the advantage $\varepsilon$. The game proceeds as follows:

As for the challenger, let $G_1$ and $G_2$ be the bilinear map with the generator $g$. Then he/she flips a coin $c$. If $c=0$, the challenger sets $<A=g^a, B=g^b, C=g^c, Z=e(g,g)^{abc}>$; otherwise, it sets $<A=g^a, B=g^b, C=g^c, Z=e(g,g)^z>$ for random $a, b, c, z$.

**Init:** The adversary $\mathcal{A}$ first selects a set of attributes $\gamma$ and send it to the challenger $\mathcal{C}$.

**Setup:** The challenger $\mathcal{C}$ runs this algorithm. The challenger $\mathcal{C}$ first sets the parameter $Y = e(A,B) = e(g,g)^{ab}$. For $\forall i \in \gamma$, it selects at random a number $r_i \in \mathbb{Z}_p$ and sets $T_i = g^{r_i}$; otherwise it chooses a random $\beta_i \in \mathbb{Z}_p$ and sets $T_i = g^{b\beta_i} = B^{\beta_i}$. Then, it gives the public parameters *PK* to the adversary $\mathcal{A}$ and keeps *MK* to itself.

**Phase 1** The adversary $\mathcal{A}$ issues queries for repeated private keys for various access structure $AS_i$. If for $\forall i, \gamma \in AS_i$, then the queries are aborted. Otherwise, $\mathcal{C}$ generates the corresponding secret keys for $\mathcal{A}$. Based on the previous definition, set the private key as:

$$D_x = \begin{cases} g^{\frac{Q_x(0)}{t_i}} = g^{\frac{bq_x(0)}{r_i}} = B^{\frac{bq_x(0)}{r_i}} & \text{if } i \in \gamma \\ g^{\frac{Q_x(0)}{t_i}} = g^{\frac{bq_x(0)}{b\beta_i}} = B^{\frac{q_x(0)}{\beta_i}} & \text{otherwise} \end{cases}$$

**Phase 2:** Repeat **Phase 1**.

**Challenge** The adversary submits two messages with equal lengths, $M_0$ and $M_1$. The challenger $\mathcal{C}$ randomly flips a coin $b$, and encrypts $M_b$ with the selected set $\gamma$. Then, the generated ciphertext $E = (\gamma, E' = m_b Z, \{E_i = C^{r_i}\}_{i \in \gamma})$ will be given to $\mathcal{A}$.

**Guess** The adversary will submit a guess $b'$ of $b$. If $b'=b$ the challenger will output $u'=0$ to indicate that it was a BDH tuple, otherwise it was a random tuple. In the case where $u=1$, the adversary gains no information about $b$. Therefore,

$$\Pr[b \neq b' | u=1] = 1/2, \Pr[u=u' | u=1] = 1/2.$$

If $u=0$, the adversary sees an encryption of $M_b$. The advantage is $\varepsilon$ by definition. Therefore,

$$\Pr[b=b' | u=0] = 1/2 + \varepsilon, \Pr[u=u' | u=0] = 1/2 + \varepsilon.$$

Thus, the overall advantage of the challenger in the DBDH game is,

$$1/2 \Pr[u=u' | u=0] + 1/2 \Pr[u=u' | u=1] - 1/2 = 1/2 \varepsilon.$$

### B. Security against Ciphertext Confusion

For reason of individual benefits, some AOs intend to collect a few ciphertexts from different areas or different communities, which they may share some of the same attributes. The half-decrypted ciphertexts will be as follows:

$$M_i e(g,g)^{ys}, C_{u,i} = T_u^{h_i(CID_i)} {}_{u \in \gamma}.$$

However, they cannot gain any information even though they perform the polynomial interpolation, where

$$g^{t_{u,i} \sum_{i=1}^{k}(h_i(CID_i) \prod_{j=1, j \neq i}^{k} \frac{CID_j}{CID_j - CID_i})} = g^{t_{u,i}s'}.$$

It is difficult to get the right $s$, therefore they cannot gain more information by confusing the ciphertexts.

### C. Selection of Participants

In both Shamir's scheme [26] and Pedersen's scheme [30], note that $2k-1 \leq n$, as the $(k, n)$ threshold schemes allow at most $k$-1 cheating participants. This means that a majority of the participants are assumed to be honest. However, in this paper, all of the RUs prefer to be able to upload their bid information accurately, as this concerns individual benefits. It is profitable for none in cheating each other.

Therefore, we relax the restriction of $k$; its value is determined by the actual number of RUs that upload information.

## V. PERFORMANCE EVALUATION

### A. Numerical analysis

In this paper, not all of the RUs need to to submit the ciphertexts at the same time, but some RUs need to submit ciphertext first, while the other ciphertexts can be delayed. Furthermore, there is not a central trusted server waiting for a long time to collect all these ciphertexts. As a result, we shorten the time that the AOs wait for decryption, and they do not need to decrypt repeatedly to get the ciphertext. Now, we give the efficiency analysis according to the encryption/decryption process. The time cost in each step is shown in Table I.

Generally, there are two intuitive schemes for RUs to upload their information to the network. One is to set a trusted CA, and it will set a transaction period $TP$. During that period of time, the RUs can upload their information independently. After time is over, the CA no longer collects information, and it will package all of the information, encrypt them together, and send the ciphertext to the network. Assume that in $TP$, there are $t$ RUs uploading their information, and for ease of description, we set each of the information size is $m$, the encryption time for all of this information is $ET_{t \cdot m}$, and the total encryption/decryption time are $T_1$, $T_1'$. Thus, the results are,

$$T_1 = TP + ET_{t \cdot m} \quad (23)$$
$$T_1' = DT_{t \cdot m} \quad (24)$$

TABLE I.    THE PARAMETERS IN PERFORMANCE EVALUATION

| Symbols | Description |
|---|---|
| $n$ | Total number of RUs |
| $TP$ | A transaction period that set by system/CA. |
| $t$ | Total number of users who upload their information in $TP$. |
| $ET$ | Encryption time of data with KP-ABE. |
| $DT$ | Decryption time of data with KP-ABE. |
| $ST$ | Time for performing the secret sharing scheme. |
| $DST$ | Time for solving the secret sharing polynomials. |
| $T$ | Total time for encryption. |
| $T'$ | Total time for decryption. |

To abolish the trusted CA, RUs only need to encrypt their information independently and then send the ciphertexts to network. Similarly, the system can set a transaction period $TP$ for $RU$s. Once the time is over, their submitted information is considered as invalid. Thus, the total encryption time $T_2$ is approximately equal to $TP$.

$$T_2 = TP \quad (25)$$

If there are $t$ RUs uploading their information, each of the ciphertexts size is $m$. This means that, in order to obtain all of the bid information, the AOs have to repeat the decryption $t$ times. Thus, the total decryption time $T_2'$ is that:

$$T_2' = t \cdot DT_m \quad (26)$$

What we can conclude from those schemes is as follows:

For Scheme 1, only when the encryption is done completely by the CA, can AOs send requests for decryption. Obviously, this scheme requires a trusted CA and results in a long waiting time for AOs.

For Scheme 2, without a CA being responsible for collecting and encrypting the RUs' information, the RUs perform the encryption independently. But this means that the AOs have to decrypt repeatedly to obtain all the information.

In this paper, we realize a flexible access control scheme that reduces the operators' waiting time for decryption. There will be a pre-specified threshold value that is set by the system according to the actual needs. Assume that in the transaction period $TP$, there are $k$ RUs generating their information and preparing to upload. They first perform the secret sharing scheme, the time we denote as $ST_{RUs}$, and then the encryption time for a single $RU$ is $ET_m$; the time that CA solves the secret sharing polynomials is denoted as $DST_{CA}$. $TP_k$ denotes the time that $k$ RUs prepares to upload their ciphertexts. Note that $TP_k < TP$. In the remaining time, other users can continue to submit information. The encryption time of $k$ RUs is,

$$T_3 = TP_k + ST_{RUs} + ET_m + DST_{CA} \quad (27)$$

When AOs require decryption, they will first solve the secret sharing polynomials, and then decrypt the ciphertexts. The decryption time is,

$$T_3' = DST_{CD} + DT_{k \cdot m} \qquad (28)$$

Compared with those two schemes, our scheme introduces a semi-trusted CA to assist with encryption. Without adding too much computational overhead, we shorten the waiting time for AO that allows them to request for decryption after $T_3(T_3 < TP)$, which does not affect the decryption of subsequent ciphertexts generated in $TP - TP_k$.

## B. Experimental results

According to the numerical analysis, without the consideration of *TP*, we first only compare the encryption/decryption time of our scheme with those two traditional schemes mentioned above, which are denoted as Scheme 1 and Scheme 2.

In the first two experiments, assume that the total number of RUs is 100 and the size of bid information is 100KB. As the number of RUs grows, this experiment primarily shows the time consuming comparison among these tree schemes.

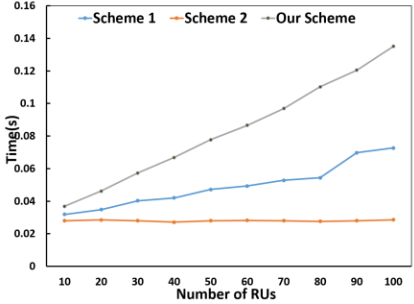
Fig.3a Comparison of the encryption time among Scheme 1, Scheme 2 and our Scheme, when the number of RUs rises.

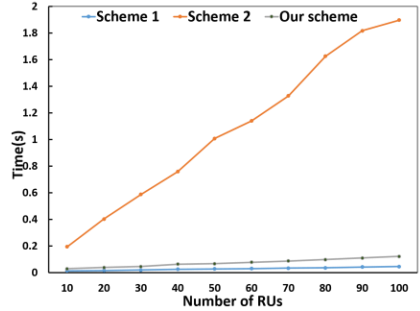
Fig.3b Comparison of the decryption time among Scheme 1, Scheme 2 and our Scheme, when the number of RUs rises.

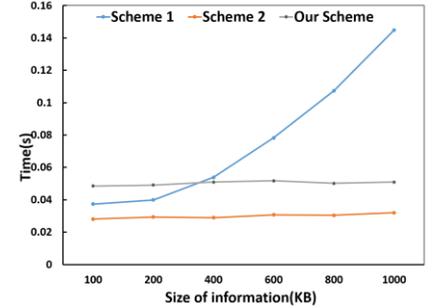
Fig.3c Comparison of the encryption time among Scheme 1, Scheme 2 and our Scheme, when the size of information rises.

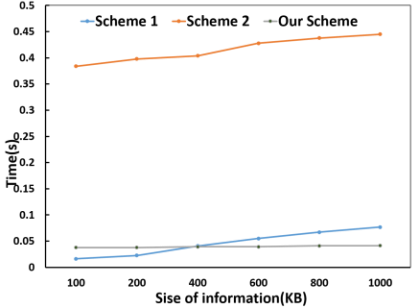
Fig.3d Comparison of the decryption time among Scheme 1, Scheme 2 and our Scheme, when the size of information rises.

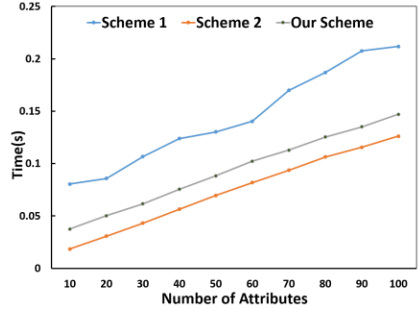
Fig.3e Comparison of the encryption time among Scheme 1, Scheme 2 and our Scheme, when the number of attributes rises.

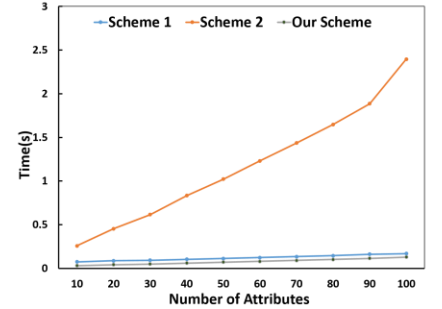
Fig.3f Comparison of the encryption time among Scheme 1, Scheme 2 and our Scheme, when the number of attributes rises.

From Fig.3a, we can find that, as the number of RUs rises, the time overhead of Scheme 1 and our Scheme is growing linearly, while Scheme 2 is approximately flat. In Scheme 1, a trusted CA is in charge of encrypting all of the RUs' information; this means that the amount of information that needs to be encrypted increases with the number of RUs. In Scheme 1, the RUs only encrypt their information independently; they are not affected by the number of RUs. In our scheme, most of the computational overhead arises on the CA's side, as it is responsible for solving the secret sharing polynomials. Although the graph shows a large amount of time overhead, when compared with the transaction period, it significantly shortens the AOs' waiting time.

In Fig.3b, unlike in the case of encryption, the time overhead of Scheme 2 is obviously higher than ours and Scheme 1. As Scheme 2 requires AOs to decrypt the ciphertexts repeatedly, the computation overhead grows linearly with the number of RUs. Similarly, the computation overhead of Scheme 1 and our Scheme also grows slightly linearly with number of RUs. In Scheme 1, the ciphertext size is linearly correlated with the number of RUs. In our scheme, a polynomial of the highest power related to the number of RUs needs to be solved first. But, the growth is still significantly lower than that of Scheme 2. Both of the two schemes require only one decryption by the AOs.

In the next two schemes, let the number of RUs be 20. As the size of information grows, this experiment mainly shows the time consuming comparison among these three schemes.

Then, we consider how the time changes when the size of the information grows. The experiment result is shown in Fig.3.c. The time overhead of Scheme 1 is most affected by the size of information, which is similar to the situation of Fig.3.a. Compared with Scheme 2, the time overhead of ours is greater, as our scheme first requires the construction of the secret sharing scheme.

However, the difference is significant during decryption showed in Fig.3d. The computational overhead of our Scheme is closer to that of Scheme 1. As in Scheme 2, AOs have to decrypt the ciphertexts repeatedly to obtain all of the information. But, in Scheme 1 and our Scheme, the decryption will be performed by the AOs only once.

Another factor affecting encryption time is the number of attributes. In the last two experiments, we let the size of information be 1MB and the number of RUs be 20. As the number of the attributes grows, this experiment primarily shows the time consuming comparison among these three schemes.

Compared to the experiments before, we can find that they are similar to each other. From Fig.3.e, it shows that all of the computation overheads of these three schemes grow linearly, while our scheme and Scheme 2 are obviously superior to Scheme 1.

In Fig.3f, obviously, the computation overhead of our scheme is closer to that of scheme 1. Repeated decryption operations take up a large amount of time in Scheme 2. But, in Scheme 1 and our scheme, the decryption will be performed by the AOs only once.

What we can draw from the above experimental results is that, in terms of computational overhead, we make a compromise between Scheme 1 and Scheme 2. In both encryption and decryption, our scheme shows a relatively lower computation overhead. Additionally, with the consideration of the previous numerical analysis, the proposed scheme shortens the time for the AOs to request and decrypt efficiently.

## VI. Conclusion and Future Work

In a scenario where the RUs upload the individual trading and bid information in the Smart Grid with RERs, we analyze the challenges regarding the security and system efficiency, and propose a flexible access control scheme with data delay tolerance for the Smart Grid with RERs. The proposed scheme makes it possible that, even within a fixed transaction period, the AOs can still access parts of the ciphertexts instead of waiting until the period is over. This will not affect the decryption of the remaining ciphertext. No matter how many ciphertexts the AOs get from the same smart community, they only need to be decrypted once. We present the security analysis to demonstrate that the proposed scheme meets the security requirements. In the performance evaluation section, we compare two other schemes with ours, which shows that the proposed scheme is efficient and practical.

In the future work, first, we will investigate how to further improve the system efficiency. Due to the limitation of traditional KP-ABE [5], we will try to realize the attribute revocation or integrate CP-ABE [21] and Proxy Re-encryption Encryption in the following scheme. Second, it is an interesting future direction to study if the proposed scheme could be integrated into some unified framework in Smart Grid (e.g. A Unified Framework for Secured Energy Resource Management in Smart Grid [15]).

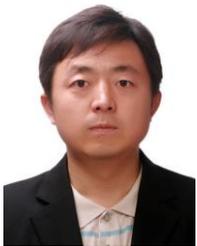

Dr. Zhitao Guan (M'13) is currently an Associate Professor in the School of Control and Computer Engineering, North China Electric Power University. He received his BEng degree and PhD in Computer Application from Beijing Institute of Technology, China, in 2002 and 2008, respectively. His current research focuses on smart grid security, wireless security and cloud security. Dr. Guan has authored over 30 peer-reviewed journal and conference papers in these areas. He is a Member of the IEEE.

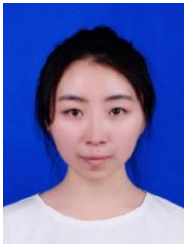

Jing Li is currently a Master candidate in the School of Control and Computer Engineering, North China Electric Power University (NCEPU). She received his BEng degree from North China Electric Power University in 2014. Her current research focuses on cloud security and smart grid security.

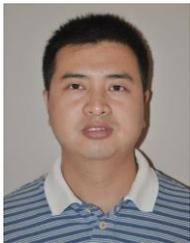

Dr. Liehuang Zhu (M') is a professor in the Department of Computer Science at Beijing Institute of Technology. He is selected into the Program for New Century Excellent Talents in University from Ministry of Education, P.R. China. His research interests include Internet of Things, Cloud Computing Security, Internet and Mobile Security.

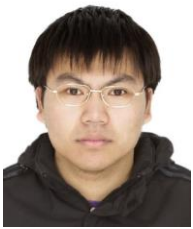

Dr. Zijian Zhang (M'13) is an assistant professor in the Department of Computer Science at Beijing Institute of Technology. He was a visiting scholar in the Computer Science and Engineering Department of the State University of New York at Buffalo in 2015. His research interests include Smart Grid, Data Privacy and Mobile Security.

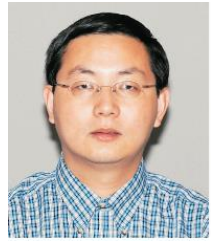

Dr. Xiaojiang (James) Du (M'04-SM'09) is currently a professor in the Department of Computer and Information Sciences at Temple University. Dr. Du received his B.S. and M.S. degree in electrical engineering from Tsinghua University, Beijing, China in 1996 and 1998, respectively. He received his M.S. and Ph.D. degree in electrical engineering from the University of Maryland College Park in 2002 and 2003, respectively. Dr. Du was an Assistant Professor in the Department of Computer Science at North Dakota State University between August 2004 and July 2009, where he received the Excellence in Research Award in May 2009. His research interests are security, wireless networks, computer networks and systems. He has published over 200 journal and conference papers in these areas. Dr. Du is a Senior Member of IEEE and a Life Member of ACM.

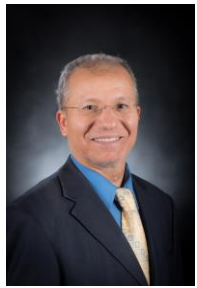

**Mohsen Guizani** (S'85-M'89–SM'99–F'09) is a Fellow of IEEE and is currently a Professor and Chair of the Electrical and Computer Engineering at the University of Idaho, USA. Previously, he served as Associate Vice President of Qatar University, Chair of Computer Science Department at Western Michigan University, Chair of Computer Science Department at the University of West Florida and Director of graduate studies at the University of Missouri-Columbia. He received his B.S. (with distinction) and M.S. degrees in Electrical Engineering; M.S. and Ph.D. degrees in Computer Engineering in 1984, 1986, 1987, and 1990, respectively, from Syracuse University. His research interests include Wireless Communications and Mobile Computing, Vehicular Communications, Smart Grid, Cloud Computing and Security. His research interests include Wireless Communications and Mobile Computing, Smart Grid, Cloud Computing and Security. He is the author/co-author of nine books and more than 450 publications in refereed journals and conferences.